\documentclass[pra,nobalancelastpage,onecolumn,nofootinbib,superscriptaddress,longbibliography]{revtex4-1}
\pdfoutput=1

\usepackage{color,amsthm,amsmath,amsxtra,amssymb , amsfonts,dsfont,graphicx,bm,bbold}
\usepackage[T1]{fontenc}
\usepackage{booktabs}

\newcommand{\be}{\begin{equation}}
\newcommand{\ee}{\end{equation}}
\newcommand{\bea}{\begin{eqnarray}}
\newcommand{\eea}{\end{eqnarray}}
\newcommand{\dpi}[2]{2\pi \times #1\,\text{#2}}
\newcommand{\gr}{g^{(2)}_{rr}}

\newcommand{\st}[1]{_\text{#1}}

\newcommand\ah{\hat a}
\newcommand\bh{\hat b}

\newcommand\sh{\hat \sigma}
\newcommand\rh{\hat \rho}
\newcommand\dg{^\dagger}
\newcommand\lind{\mathcal{L}}

\newcommand\hc{\text{h.c.}}

\newcommand{\ev}[1]{\left\langle #1\right \rangle}
\newcommand{\bra}[1]{\left\langle #1\right|}
\newcommand{\ket}[1]{\left| #1\right\rangle}
\newcommand{\lla}[1]{\left\{ #1\right \} }

\newcommand{\braket}[2]{\langle #1|#2\rangle}

\newcommand{\pa}[1]{\left( #1\right)}

\newcommand{\abs}[1]{\left| #1\right|}

\newcommand{\ham}[1]{\hat H_{\text{#1}}}

\def\multiset#1#2{\ensuremath{\left(\kern-.3em\left(\genfrac{}{}{0pt}{}{#1}{#2}\right)\kern-.3em\right)}}

\usepackage{accents}

\usepackage{hyperref}

\definecolor{mygrey}{gray}{0.35}
\definecolor{myblue}{rgb}{0.2,0.2,0.8}
\definecolor{myzard}{cmyk}{0,0,0.05,0}
\definecolor{mywhite}{rgb}{1,1,1}
\definecolor{mywhite}{rgb}{1,1,1}
\definecolor{myred}{rgb}{1,0.,0.3}

\newcommand{\co}[1]{\left[ #1\right]}
\newcommand{\tr}[1]{\text{Tr}\pa{#1}}

\hypersetup{
  pdfsubject={},
  pdfcreator={pdflatex},
  colorlinks,
  linkcolor=blue,
  citecolor=blue,
  urlcolor=blue
}
\newcommand{\ICFO}{ICFO-Institut de Ciencies Fotoniques, The Barcelona Institute of Science and Technology, 08860, Castelldefels (Barcelona), Spain}

\begin{document}
\title{Optomechanical strong coupling between a single cavity photon and a single atom
}
\date{\today }

\begin{abstract}
Single atoms coupled to a cavity offer unique opportunities as quantum optomechanical devices because of their small mass and strong interaction with light. A particular regime of interest in optomechanics is that of "single-photon strong coupling," where motional displacements on the order of the zero-point uncertainty are sufficient to shift the cavity resonance frequency by more than its linewidth. In many cavity QED platforms, however, this is unfeasible due to the large cavity linewidth. Here, we propose an alternative route in such systems, which instead relies on the coupling of atomic motion to the much narrower cavity-dressed atomic resonance frequency. We discuss and optimize the conditions in which the scattering properties of single photons from the atom-cavity system become highly entangled with the atomic motional wave function. We also analyze the prominent observable features of this optomechanical strong coupling, which include a per-photon motional heating that is significantly larger than the single-photon recoil energy, as well as mechanically-induced oscillations in time of the second-order correlation function of the emitted light. This physics should be realizable in current experimental setups, such as trapped atoms coupled to photonic crystal cavities, and more broadly opens the door to realizing qualitatively different phenomena beyond what has been observed in optomechanical systems thus far.
\end{abstract}

\author{Javier Arg\"uello-Luengo}
\affiliation{\ICFO}
\email{javier.arguello@icfo.eu}
\author{Darrick E. Chang}
\affiliation{\ICFO}
\affiliation{ICREA-Instituci\'o Catalana de Recerca i Estudis Avan\c{c}ats, 08015 Barcelona, Spain}

{
\let\clearpage\relax
\maketitle
}
Quantum optomechanics has emerged as a field with numerous exciting prospects for fundamental science and applications~\cite{Aspelmeyer2014,Metcalfe2014}. Generically, such systems are characterized by some mechanical degree of freedom, whose small displacements alter the resonance frequency of a cavity. This results in rich backaction effects once the cavity is driven that include sensing~\cite{Schreppler2014,Wu2017,Gil-Santos2020,Fischer2019}, cooling of the mechanical mode~\cite{Chan2011,Teufel2011}, generation of squeezed light~\cite{Purdy2014,Safavi-Naeini2013,Aggarwal2020} or the creation of nonreciprocal devices~\cite{Xu2019}.
A key figure of merit is the vacuum optomechanical coupling strength, $g_0=(\partial\omega_c/\partial x)\, x\st{zp}$, given by the product of the sensitivity of the cavity frequency to position displacements, and the zero-point motion of the resonator. In particular, the single-photon, single-phonon, strong coupling regime ensues when $g_0$ exceeds the linewidth of the cavity, such that the optical response and dynamics change drastically at the level of individual quanta. For example, it has been proposed that this can give rise to quantum optical nonlinearities~\cite{Rabl2011,Nunnenkamp2011}.
While a number of schemes have been proposed to reach this strong coupling regime~\cite{Brunelli2020,Yin2017,Lemonde2016,Pirkkalainen2015a,Heikkila2014}, state-of-the-art experimental setups remain at least two orders of magnitude away from reaching this regime~\cite{MacCabe2020,ren2020two}. 

Here, we show that the single-photon strong coupling regime can be realistically achieved using a single atom coupled to a high-finesse cavity~\cite{Shomroni2014,Birnbaum2005a,Reiserer2015a,Hacker2019,Hamsen2018}. A single atom constitutes an interesting candidate for an optomechanical element, due to its low mass and anomalously large optical response (i.e. a scattering cross section much larger than its physical size). While macroscopic cavity architectures allow for sufficiently small linewidths to reach the strong coupling regime~\cite{Hamsen2018,Brennecke2008,Birnbaum2005a,Purdy2010,Neumeier2018b,Neumeier2018}, a number of platforms~\cite{Tiecke2014,Thompson2013a,Samutpraphoot2019,Bechler2018,Will2021} focus on achieving small mode volumes with a prohibitively large linewidth.
Here, we show that optomechanical strong coupling effects can nonetheless emerge in these devices by working in a detuned atom-cavity regime and probing motional interactions on the narrower dressed atomic resonance. 

The enabling mechanism is based on the scattering properties of an incoming photon, which highly depend on its detuning to the dressed resonance frequency that in turn is sensitive to the atomic position within the cavity field. As a consequence, we show that a scattered photon highly entangles with the resulting atomic motional state, carrying information about its position. This leads to to a per-photon atomic heating larger than expected from single-photon recoil events and, as a more direct signature, we observe that detection of a reflected photon triggers motion-induced oscillations in time of the second-order correlation function of reflected light. We show that these effects are observable in realistic systems, even for a non-zero initial motional temperature.

\subsection{The system}
\begin{figure*}[tbp]
\centering
  \includegraphics[width=0.65\linewidth]{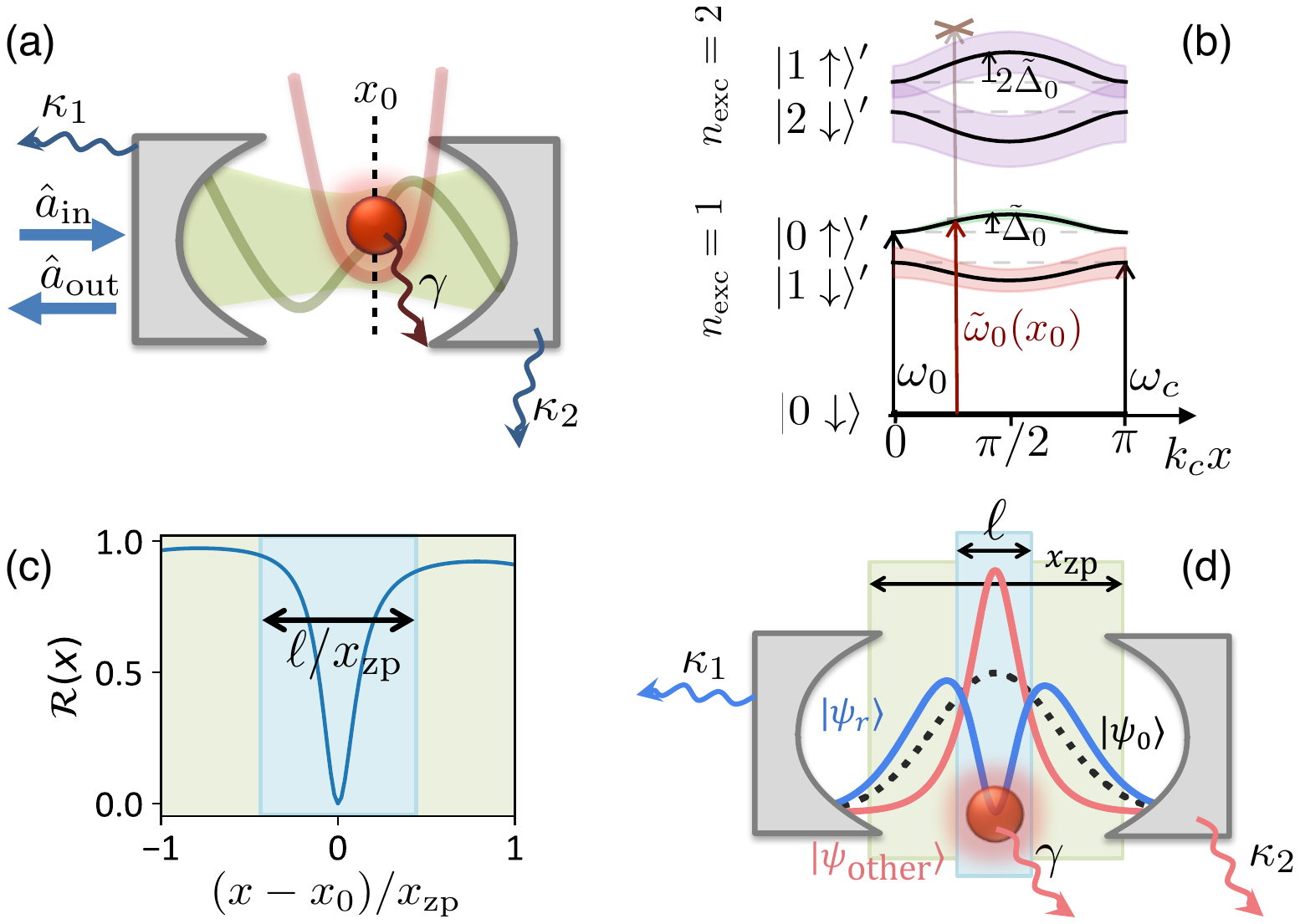}
    \caption{(a) A single atom is trapped in a harmonic oscillator potential of frequency $\omega_m$ centered at position $x_0$, and coupled to a cavity mode. The cavity mode is externally driven with input field $\ah\st{in}$ and can decay through this driving channel ($\kappa_1$), or other undetected routes ($\kappa_2$). In addition, the excited atom can spontaneously emit into free space at a rate $\gamma$. (b) Schematic representations of the lowest energy levels of the J-C Hamiltonian in the absence of a drive (analogous to considering $\varepsilon=0$ and $\omega_l=0$ in~\eqref{eq:hamJC}), showing up to 2 total excitations. For atomic positions away from the cavity nodes ($k_cx=0,\pi$), the frequencies of the dressed eigenstates $\ket{n\st{ph},\uparrow/\downarrow}'$ experience a shift from the uncoupled levels $\ket{n\st{ph},\uparrow/\downarrow}$ that depends on atomic position, $\tilde \Delta_0\approx g^2(x)/\Delta$. The linewidth of these dressed levels is represented by shaded regions in the situation $\gamma\ll \kappa $ explored in this work. (c) In the studied configuration, reflectance is tuned to be null when the atom is placed at the center of the trap $x_0$, with the spatial width of this reflection minimum given by $\ell$.
  (d) After an incoming photon is scattered, the initial motional wave function $|\psi_0\rangle $ is strongly modified over the length $\ell$, to the state $|\psi_r\rangle$ conditioned on the reflection of a photon, or the state $|\psi\st{other}\rangle$ conditioned on scattering into other channels.}
 \label{fig:schemeLoss}
\end{figure*}

Here we focus on the interaction of a single two-level atom with an optical transition between ground and excited states $\ket{\downarrow}$, $\ket{\uparrow}$, and a given mode of the electromagnetic field inside the cavity. 
The coherent interactions are described by the Jaynes-Cummings (J-C) Hamiltonian~\cite{Jaynes1963} for a single atom,
\begin{align}
    \begin{split}
\label{eq:hamJC}
    \ham{JC}=&-(\Delta+\Delta_0) \ah\dg \ah  -\Delta_0 \sh\dg \sh \\
    &+ g(\hat x)\pa{\sh \ah\dg +\hc}-i\varepsilon\pa{\ah\dg-\ah}\,,
    \end{split}
\end{align}
where $\sh\equiv \ket{\downarrow}\bra{\uparrow}$ is the atomic lowering operator, and $\Delta=\omega_0-\omega_c$ the energy difference between the bare atomic and cavity resonance frequencies. Here, we also allow for an external laser drive of the cavity with $\Delta_0=\omega_l-\omega_0$ representing the laser-atom detuning, and $\varepsilon$ the driving amplitude, which we will generally consider weak enough to only produce a few excitations.
$g(\hat x)=g_0\sin(k_c\hat x)$ denotes the position-dependent vacuum Rabi coupling strength, with $k_c$ being the cavity mode wavevector. Importantly, we will treat the atomic position $\hat{x}$ as a quantum dynamical degree of freedom, and assume that the atom is harmonically trapped with frequency $\omega_m$ and equilibrium position $x_0$, as schematically illustrated in Fig. \ref{fig:schemeLoss}(a). One can quantize the atomic motion around this point as $\delta \hat x=\hat x-x_0=x\st{zp}(\bh+\bh^\dagger)$, where $x\st{zp}$ denotes zero-point motion fluctuations, and $\bh^{(\dagger)}$ is the anihilation(creation) operator of phonons in the trap, $ \hat H\st{trap}=\omega_m \bh^\dagger \bh$. Here and in what follows we use the convention that $\hbar\equiv 1$.

Further including photonic losses from the cavity with decay rate $\kappa$, and atomic excited state spontaneous emission at a rate $\gamma$, one can describe the total evolution of the density matrix as,
\begin{equation}
\label{eq:JCevolution}
    \frac{d\rh}{dt}=-i \co{\hat H \st{JC}+\hat H\st{trap},\rh}+\kappa\, \lind\co{\ah}\pa{\rh}+\gamma \,\lind\co{e^{-ik_c\hat x}\sh}\pa{\rh}\,,
\end{equation}
where we define the Lindbladian $\lind\co{\ah}\pa{\rh}=\frac{1}{2}\pa{2\ah \rh \ah\dg -\ah\dg\ah\rh - \rh \ah\dg\ah}$. 
The $e^{-ik_c \hat x}$ term represents the recoil momentum that is imparted onto the atom upon spontaneous emission of a photon. Strictly speaking, the recoil along the $x$ direction is a (non-uniform) random variable between $(-k_c,k_c)$, accounting for the possibility of a photon to be emitted in any direction~\cite{Cirac1992}, but the simplification made above is sufficient to capture all of the salient physics.

Here, we will consider the regime relevant to a number of cavity QED systems, where $\kappa\gg\gamma$~\cite{Bechler2018,Samutpraphoot2019,Will2021}. In order to access a strong optomechanical coupling, this motivates working in a detuned atom-cavity regime $\abs{\Delta}\gg \kappa,g_0$ and focusing on the dressed atom-like excitation branch with narrower linewidth $\sim\gamma$. We will start by presenting some heuristic arguments to estimate the optimal conditions to reach this single-photon optomechanical strong coupling, which we will later show are rigorously correct. To simplify the discussion, we will also start by considering the case where the atom is initialized in the motional ground state, treating thermal states in Section~\ref{ap:finiteTemp}.

\subsection{Heuristic derivation of strong coupling condition in the regime, $|\Delta|\gg g_0,\kappa$}
One can start by considering a static atom with fixed position $x$. 
In the absence of a drive ($\varepsilon=0$), one can block diagonalize the J-C Hamiltonian~\eqref{eq:hamJC} in the total number of excitations, $n\st{exc}\equiv \sh\dg\sh+\ah\dg\ah$, as illustrated in Fig.~\ref{fig:schemeLoss}(b) for up to $n\st{exc}=2$. In the limit of large atom-cavity detuning $\Delta$, it is well-known that one of the single-excitation eigenstates $\ket{0\uparrow}'$ is mostly an atomic excitation $\ket{0\uparrow}$, but with a shifted resonance frequency,
$
    \tilde \omega_0(x)\approx \omega_0+\frac{g^2(x)}{\Delta}\,,
$
and broadened linewidth,
$
        \tilde \gamma (x)\approx \gamma + \kappa\frac{g^2(x)}{\Delta^2}\,,
$
due to the interaction with the cavity.
We can consider the sensitivity of this resonance frequency to small (static) displacements $x=x_0+\delta x$, which to lowest order yields a new resonance frequency 
$
       \tilde \omega_0(x_0+\delta x) \approx \omega_0+\frac{g_0^2}{\Delta} \Big[ \sin^2(k_cx_0) +\sin(2k_c x_0) \, k_c  \delta x
    \Big]\,.
$
The maximum sensitivity to a displacement $\delta x$ then occurs halfway between a cavity node and anti-node, when $k_cx_0=\pi/4$ (see Fig. \ref{fig:schemeLoss}(b)). 

Although we take $x$ to be static, one can nonetheless intuitively deduce a single-photon optomechanical strong coupling parameter, 
$
    \beta \equiv \frac{g_0 ^2\eta}{\Delta\, \tilde\gamma(x_0)}\,,
$
which characterizes how much the dressed resonance frequency shifts if the atom is displaced by the zero-point motion, in units of the dressed linewidth. Optimizing over $\Delta$, one observes that the maximum strong coupling parameter is dictated by the cooperativity, $C\equiv g_0^2/(\kappa\gamma)$, as $\beta\st{max}=\eta\sqrt{C}/\sqrt{2}$, where $\eta\equiv k_c x\st{zp}$ is the Lamb-Dicke parameter.

We now derive the reflection coefficient of a weak monochromatic, coherent input field, as a function of atomic position. For this, we distinguish the decay rate of the cavity into the port used to drive the system $(\kappa_1)$, from the decay into transmission or absorption channels $(\kappa_2)$, so that the total cavity decay rate reads as $\kappa=\kappa_1+\kappa_2$ (see Fig. \ref{fig:schemeLoss}(a)). 
In particular, the input-output formalism~\cite{Gardiner1985,Gardiner2014} allows us to express the field $\ah\st{out}$ leaving the cavity through the channel associated to $\kappa_1$ in terms of the input field, $\ah\st{in}$, as
\begin{equation}
\label{eq:inputOutput}
    \ah\st{out}(t)=\ah\st{in}(t)+\sqrt{\kappa_1}\ah(t)\,,
\end{equation}
which satisfies, $\co{\ah\st{in}(t),\ah\dg\st{in}(t')}=\delta(t-t')$, and $\ah\st{in (out)}\dg \ah\st{in (out)}$ has units of photon number per unit time. 

For an atom statically located in position $x$, one can define $S_r(x)$ as the steady-state reflection coefficient defined by the ratio between output and input fields~\cite{Xu2015,caneva2015quantum},
\begin{equation}
\label{eq:reflFactorT}
    S_r(x)=\frac{\ev{\ah\st{out}}}{\ev{\ah\st{in}}}\approx 1-\frac{i\kappa_1}{\Delta+\Delta_0+i\kappa/2-\frac{g(x)^2}{\Delta_0+i\gamma/2} }\,,
\end{equation}
and a corresponding reflectance, $\mathcal R(x) =\abs{S_r(x)}^2$, as a function of the atomic position.
Intuitively, efficient optomechanical coupling requires a large contrast in $\mathcal R(x)$ when the atom is displaced from position $x_0$ by a small amount, so that the event of detecting a reflected photon reveals significant information about the atomic position. Experimentally, one can optimize this by adjusting the driving frequency, the atom-cavity detuning, and the coupling to the detection channel, $\kappa_1$. 
First, we choose to drive the atom-like resonance,
\begin{equation}
\label{eq:optoml}
    \Delta_0^*=\frac{g^2(x_0)}{\Delta^*}\,.
\end{equation}
Expanding now the reflectance of the cavity around $x_0$, $\mathcal{R}(x)=\mathcal{R}_0 + [(x-x_0)/\ell]^2$, one can enforce that the reflectance at position $x_0$ is exactly zero, $\mathcal R_0=0$, so that detection of a reflected photon ensures that the atom is not placed at that point. Imposing this, one obtains the optimal detuning,
\begin{equation}
\label{eq:optdelc}
    \Delta^*=g(x_0)\sqrt{\frac{\kappa_1-\kappa_2}{\gamma}}\,,
\end{equation}
which corresponds to critical coupling, where the (dressed) atomic excited state decays equally into the cavity output, $\pa{\kappa_1\frac{g^2(x_0)}{\Delta^2}}$, and other channels, $\pa{\gamma+\kappa_2\frac{g^2(x_0)}{\Delta^2}}$. 
Maximizing the effective single-photon coupling parameter $\beta$ as a function of $\kappa_1$ for the previous choice of parameters, one obtains
$
    \kappa_1^*=2\kappa_2\,.
    $
This in turn yields the minimum displacement, 
\begin{equation}
\label{eq:R} 
    \ell^*/x\st{zp}\equiv \sqrt{2}/\pa{\eta\sqrt{C\st{in}}}\,,
\end{equation}
over which the dressed atomic frequency shifts by $\tilde \gamma$, thus bringing the system off resonance with respect to the fixed external laser frequency (see Fig. \ref{fig:schemeLoss}(c)). The fact that $\ell$, representing the length scale over which a single photon can discriminate the atomic position, depends inversely with the square root of the intrinsic cooperativity $C\st{in}\equiv g_0^2/\sqrt{\kappa_2\gamma}$ will play a prominent role in our following discussion. In particular, we 
observe that the maximum strong coupling parameter previously defined scales as $\beta^* \sim x\st{zp}/\ell^*$.

\subsection{Role of the atomic motional wave function}
\label{sec:heating}
Previously, we have established that if the atom was a perfectly localized point particle, a photon would be reflected with an amplitude and phase given by $S_r(x)$. Intuitively, once the atomic motional state is given by a wave function $\ket{\psi_0}=\int dx\, \psi_0(x)\ket{x} $, one might expect that the state upon scattering a single photon is given by $S_r\ket{1_r}\ket{\psi_0}+S\st{other}\ket{1\st{other}}\ket{\psi_0}$, where $\ket{1_r}$ denotes a reflected photon, and $\ket{1\st{other}}$ denotes the scattering into some orthogonal channel (transmission or cavity absorption, see Appendix \ref{ap:scattering}). While $S_r\ket{1_r}\ket{\psi_0}=\ket{1_r}\int dx\, S_r(x) \psi_0(x)\ket{x}$ has the natural meaning that the amplitude and phase of the reflected photon depends on the atomic position, one can also observe that the atomic wave function conditioned on the detection of the reflected photon becomes,
\begin{equation}
\label{eq:psir}
    \ket{\psi_r}=\frac{S_r \ket{\psi_0}}{\abs{S_r \ket{\psi_0}}}\,,
\end{equation}
where $S_r \ket{\psi_0}=\int dx\, S_r(x) \psi_0(x)\ket{x}$ and the denominator relates to the average reflectance of the cavity, $\mathcal R\equiv \abs{S_r \ket{\psi_0}}^2= \int dx\, \mathcal R(x) \abs{\psi_0(x)}^2$.
These results, which were up to now argued intuitively, can in fact be derived rigorously through an adiabatic elimination of the cavity degrees of freedom in the unresolved sideband regime, $\omega_m\ll \gamma,\kappa$~\cite{Neumeier2018}, where the dynamics of the atom-cavity interface is much faster than the mechanical evolution of the atom inside the trap and $S_r$ defines a scattering matrix~\cite{Xu2015,caneva2015quantum} that is diagonal in the position basis (Appendix \ref{ap:scattering}).

Intuitively, if the reflectance of the cavity was similar for different atomic positions, $\mathcal R(x)\approx \mathcal R$, reflection would reveal no information and the wave function would remain unaffected by detection, $\ket{\psi_r}\approx \ket{\psi_0}$. In contrast, if $S_r(x)$ contains any narrow spatial features, 
those features are now imprinted onto the atomic wave function itself. 
To quantify this, 
we observe that $\ell$ sets the characteristic width for the spatial features imprinted in the wave function (see Fig.~\ref{fig:schemeLoss}(d)). For an atom initially in the ground state of the trap, the effect of detection will then be large when this critical displacement is smaller than the zero-point motion of the atom, $\ell\ll x\st{zp}$, which corresponds to the strong coupling regime, $\eta\sqrt{C\st{in}}\gg 1$.

\subsection{Unconventional heating}
As the initial atomic state $\ket{\psi_0}$ is modified by events associated to reflection or emission in other channels, its mechanical energy departs from the trap ground state energy. Following the example of Eq. \eqref{eq:psir}, one can calculate the average number of phonons induced by a single incident photon, $\mathcal{J}=\mathcal{J}_r+\mathcal{J}_t+\mathcal{J}_a$, associated to scattering in the detection channel, cavity transmission/absorption, or atomic spontaneous emission, respectively, where
\begin{equation}
\label{eq:tildeH}
    \mathcal{J}_\alpha=\bra{S_\alpha\psi_0}\bh\dg \bh \ket{S_\alpha \psi_0}
\end{equation}
for $\alpha\in\lla{r,t,a}$.
Note that for each of these emission mechanisms, $\mathcal{J}_\alpha$ then represents the number of phonons in the resulting atomic state $\bra{S_\alpha\psi_0}\bh\dg\bh\ket{S_\alpha \psi_0}/\braket{S_\alpha\psi_0}{S_\alpha \psi_0}$, weighted by the probability that this event occurs, $ \braket{S_\alpha\psi_0}{S_\alpha \psi_0}$ (see Appendix \ref{ap:scattering}).

 \begin{figure}[tbp]
\centering
\includegraphics[width=0.65\linewidth]{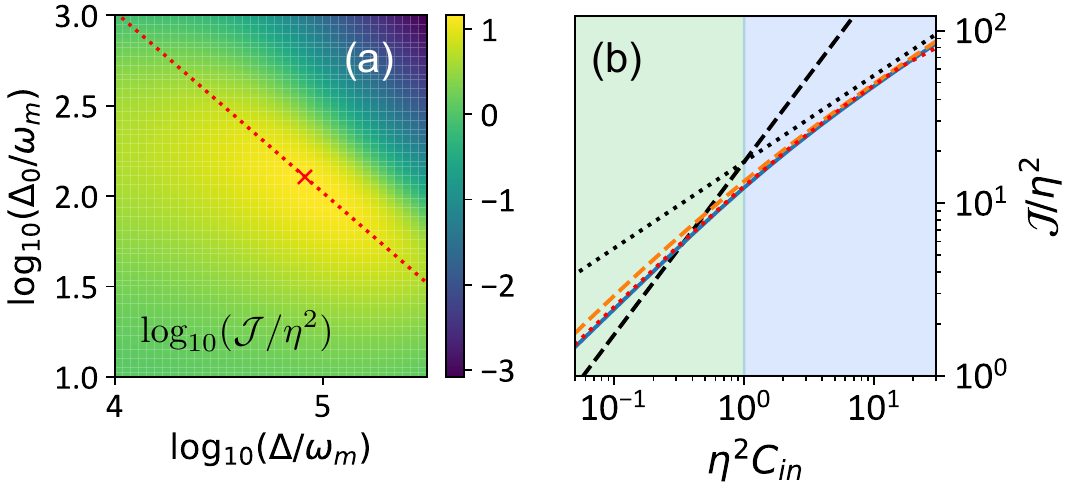}
  \caption{(a) Ratio of the per-photon increase in phonons ($\mathcal J$) of an atom in the cavity compared to the free-space result $\eta^2$,  for different choices of $\Delta$ and $\Delta_0$. Here $\kappa_1=2\kappa_2$ and $k_c x_0=\pi/4$ (see main text). When not indicated otherwise, parameters compatible with Ref.~\cite{Samutpraphoot2019} are used in the Figures: $ g_0=\dpi{0.73}{GHz},\,\omega_m=\dpi{160}{kHz},\,\gamma=\dpi{6}{MHz},\,\kappa_2=\dpi{3.9}{GHz},\,\eta=0.24$. The red dotted line follows a resonant driving with the dressed atomic frequency \eqref{eq:optoml}, and the crossed marker indicates the optimal atom-cavity detuning \eqref{eq:optdelc}.
(b) Value of $\mathcal{J}/\eta^2$, maximized over free values of $\Delta$ and $\Delta_0$ (orange dashed line), as compared to the result associated to $\Delta^*,\, \Delta_0^*$ within our effective model (blue line), for increasing intrinsic cooperativity tuned by varying $g_0$. The red dotted line corresponds to a master equation simulation of the open system (see main text). Dashed and dotted black lines follow the scalings $\mathcal J\sim \eta^2 C\st{in}$ and $\mathcal J \sim \eta \sqrt{C\st{in}}$ expected in the regimes $\eta^2 C\st{in}\ll 1$ (coloured in green), and $\eta^2C\st{in}\gg 1$ (coloured in blue), respectively.}
 \label{fig:60MechanicalProjectionScattering_heatingScalingConditions_dat}
\end{figure}

For conventional scattering from a tightly trapped atom in free space, the characteristic number of phonons that an incoming photon can excite is characterized by the ratio between the single-photon recoil energy, $\omega_r$, and the mechanical frequency of the oscillator. When expressed in terms of the Lamb-Dicke parameter, this translates to a per-photon increase in phonons of $\eta^2=\omega_r/\omega_m$~\cite{Gardiner2014}.

In our coupled atom-cavity system, this heating effect can now be enhanced.
Based on our previous analysis,
we expect that the largest values of $\mathcal J$ will appear for the choices of cavity-laser and atom-cavity detunings derived in Eqs. (\ref{eq:optoml},\ref{eq:optdelc}). 
To validate this, in Fig. \ref{fig:60MechanicalProjectionScattering_heatingScalingConditions_dat}(a) we numerically calculate $\mathcal{J}/\eta^2$ for different detunings $\Delta_0$ and $\Delta$
in a cavity satisfying $\kappa_1=2\kappa_2$. The rest of the parameters are compatible with current experimental platforms~\cite{Samutpraphoot2019} where one can reach large intrinsic cooperativities in the order of $C\st{in}\sim 23$, and a Lamb-Dicke parameter $\eta\sim 0.24$.
In agreement with our derivation, we observe that a driving frequency in resonance with the dressed atomic frequency (red dotted line, Eq. \eqref{eq:optoml}) corresponds to the region of larger heating ($\mathcal J\sim 10\eta^2$, associated to lighter colors), and that the atom-cavity detuning that produces maximal heating is compatible with the prediction of Eq. \eqref{eq:optdelc} (crossed marker).

In Fig. \ref{fig:60MechanicalProjectionScattering_heatingScalingConditions_dat}(b) we calculate $\mathcal{J}/\eta^2$ for the choices of $\Delta_0$ and $\Delta$ predicted to maximize optomechanical coupling (Eqs. (\ref{eq:optoml},\ref{eq:optdelc})), but now as a function of the intrinsic cooperativity $C\st{in}$ by allowing the vacuum Rabi coupling $g_0$ to vary, while maintaining the rest of the experimental parameters ($\omega_m$, $\gamma$, $\kappa_2$, $\eta$) as before. We observe that, for each value of intrinsic cooperativity, the heating that arises at the optimal parameters $\Delta_0^*$ and $\Delta^*$ (blue line) does match a fully numerical maximization of the average number of induced phonons over free values of $\Delta_0,\, \Delta$ (orange dashed line), quantitatively confirming our analysis.

While these calculations were based on the scattering matrix given in Eq.~\eqref{eq:reflFactorT} and Appendix~\ref{ap:scattering}, we additionally validate these results by performing a master equation simulation of the open system (red dotted line) in a truncated space of up to 2 photons and 50 phonons, where convergence is observed. For this, we evolve under Eq.~\eqref{eq:JCevolution} a density matrix with initially no excitations in the system (cavity photons, atomic excitations or phonons) until the population of the cavity stabilizes (variations smaller than 1$\%$). Using input-ouput relations
analogous to Eq.~\eqref{eq:inputOutput} and normalizing by the number of
incoming photons, the three heating contributions can be calculated as $\mathcal{J}_{r/t}=\kappa_{1/2}\tr{\bh\dg\bh \, \ah\rh\st{ss}\ah\dg }/\pa{\varepsilon^2/\kappa_1}$, and $\mathcal{J}_{a}= \gamma \tr{\bh\dg\bh\, \sh\rh\st{ss}\sh\dg }/\pa{\varepsilon^2/\kappa_1}$. Adding them up (red dotted line), we observe in Fig. \ref{fig:60MechanicalProjectionScattering_heatingScalingConditions_dat}(b) good agreement with the scattering matrix calculation for the considered weak driving amplitude $\varepsilon^2/\kappa_1= 0.01\,$ MHz.

Interestingly, we observe that the average number of induced phonons $\mathcal{J}$ scales differently with the intrinsic cooperativity in the weak ($\eta^2 C\st{in}\ll 1$) and strong coupling regimes ($\eta^2 C\st{in}\gg 1$), which can be understood from the phase $\varphi_\alpha(x)$ of the scattering matrices $S_\alpha \sim e^{i\varphi_\alpha(x)}$ that gets imprinted onto the atomic wave function.
 For the optimal parameters (\ref{eq:optoml},\ref{eq:optdelc}), we note that up to linear order in $\delta x/x\st{zp}$ the imprinted phase associated to emission in undriven channels scales linearly as $\varphi_{t/a}(x)\sim \eta\sqrt{C\st{in}}\delta x/x\st{zp}$ (see Eq. \eqref{eq:delsig}), which corresponds to an added momentum of $\eta\sqrt{C\st{in}}/x\st{zp}=\sqrt{2}/\ell$.
Although $S_r(x)$ cannot be expressed as a phase term, atomic heating can only depend on the total cavity decay rate $\kappa$, and not on the specific channel contributing to this rate. Thus, as $\kappa_1=2\kappa_2$, it follows that the heating rate due to reflection is twice that of transmission/absorption, $\mathcal J_r\approx 2\mathcal J_t$. 

Adding these three contributions in the weak-coupling limit, ($x\st{zp}\ll \ell$, green shaded region of Fig.~\ref{fig:60MechanicalProjectionScattering_heatingScalingConditions_dat}(b)), the imprinted momentum affects the entire wave function and the corresponding kinetic energy increase leads to a heating rate of $\mathcal{J}^{\ll} \sim (x\st{zp}/\ell)^{2} \sim \eta^2 C\st{in}$. In contrast, in the strong optomechanical coupling limit ($x\st{zp}\gg \ell$, blue shaded region), the phase imprinting only applies to a small region $\ell$ of the entire wave function, where the cavity is actually sensitive to the atomic position. This leads to a heating rate of $\mathcal{J}^{\gg}  \sim (x\st{zp}/\ell)^2 \cdot (\ell/x\st{zp}) \sim \eta \sqrt{C\st{in}}$, matching the scalings observed in Fig. \ref{fig:60MechanicalProjectionScattering_heatingScalingConditions_dat}(b) (black dashed and dotted lines, respectively). 
The fact that the per-photon heating rate could be one or two orders of magnitude larger than the expected free space result could be relevant to experiments that probe around the dressed atomic resonance frequency. Separately, we note that the enhanced heating of an atomic ensemble has been experimentally observed in a complementary regime, driving around the dressed cavity resonance of a detuned atom-cavity system~\cite{Murch2008}.

\begin{figure}[t]
\centering
\includegraphics[width=0.6\linewidth]{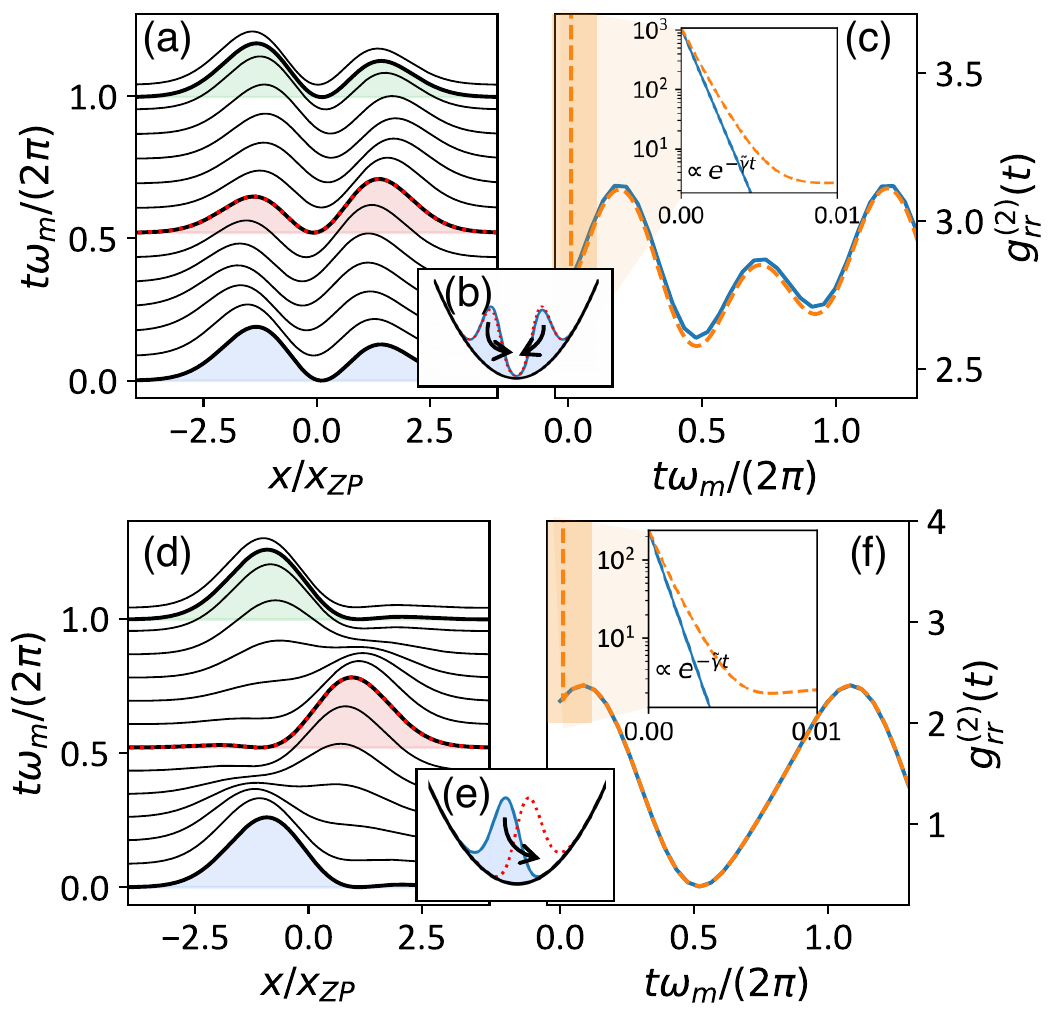}
  \caption{(a) Spatial probability distribution $\abs{\psi_r(t)}^2$ following the detection of a reflected photon at $t=0$, and under the assumption that the subsequent motional wave function only evolves under the external trapping potential for the configuration $\eta=0.05$, $\kappa_1/\kappa_2=1.6$, marked with a red cross in Fig. \ref{fig:71optimizingG2_k1k2}(a). Later times are indicated by vertical shifts of the atomic density, and the wave function at times $\pi/\omega_m$ and $2\pi/\omega_m$ are coloured in red and green, respectively. Following Eqs. (\ref{eq:optoml}, \ref{eq:optdelc}), $\Delta_0$ and $\Delta$ are chosen to satisfy $\mathcal R(x_0)=0$. Rest of parameters as in Fig. \ref{fig:60MechanicalProjectionScattering_heatingScalingConditions_dat}. (b) Schematic representation of $\abs{\psi_r}^2$ at time $t=0$ (blue) and $t=\pi/\omega_m$ (dotted red). (c) Calculation of $\gr( t)$ along this evolution, using the scattering matrix approach (blue line) and a master equation simulation of the cavity system (orange dashed line). The inset highlights the fast decay of $\gr(t)$ at initial times revealed by the master equation (dashed line), compatible with a decay rate $\ev{\tilde \gamma(x)}$ (continuous line). (d-f) Analogous plots to (a-c), now for the case $\mathcal R(x_0+x\st{zp})=0$ (see text). }
 \label{fig:44MechanicalProjectionScatteringExpEv}
\end{figure}

\begin{figure}[tbp]
\centering
  \includegraphics[width=0.6\linewidth]{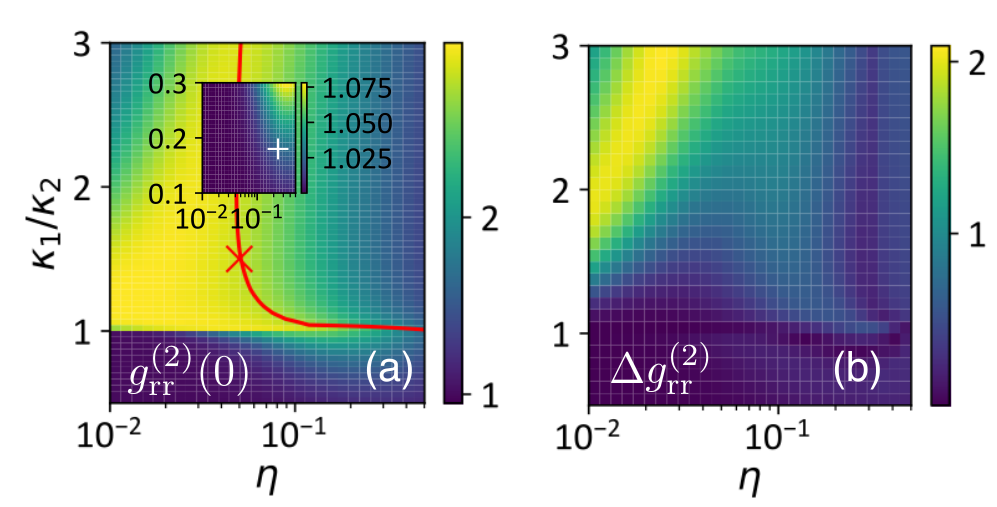}
  \caption{(a) Scattering matrix calculation of $\gr(0)$ when driving in resonance the dressed atomic frequency for the configuration $\mathcal R_0=0$ as defined in Eqs. (\ref{eq:optoml},\ref{eq:optdelc}), as one varies the Lamb-Dicke parameter $\eta$ and the ratio $\kappa_1/\kappa_2$. For $\kappa_1<\kappa_2$, where it is not possible to obtain $\mathcal R_0=0$, we numerically maximize $\gr(0)$ as a function of $\Delta$. Rest of experimental parameters as in Fig. \ref{fig:60MechanicalProjectionScattering_heatingScalingConditions_dat}. Red line follows the relation $\ell(1+\mathcal{R}_0)=6 x\st{zp}$, and red marker indicates the configuration $\kappa_1/\kappa_2=1.6$, $\eta=0.05$ explored in Fig. \ref{fig:44MechanicalProjectionScatteringExpEv}. The inset zooms into the region of small ratios $\kappa_1/\kappa_2$, and the white marker signals the configuration $\kappa_1/\kappa_2=0.18$, $\eta=0.24$, compatible with \cite{Samutpraphoot2019}, where $\gr(0)\approx 1.03$. (b) For the same parameter choices as (a), we illustrate the overall variation along a full mechanical oscillation, $\Delta \gr\equiv \max_t \gr(t)-\min_t \gr(t)$. 
  }
 \label{fig:71optimizingG2_k1k2}
\end{figure}

\subsection{Second-order time correlations}
We now consider how the strong optomechanical coupling can manifest itself in the second-order time correlations of the reflected field,
\begin{equation}
\label{eq:g2out}
    \gr( t)\equiv\frac{\ev{\ah\st{out}\dg(0) \ah\st{out}\dg( t) \ah\st{out}( t) \ah\st{out}(0)}}{\ev{\ah\st{out}\dg( t)\ah\st{out}( t)} \ev{\ah\st{out}\dg(0) \ah\st{out}(0)} }\,,
\end{equation}
which quantifies the relative likelihood of detecting a reflected photon at time $t$, given the previous detection of a reflected photon at time $t=0$.

We first present an approximate theory, based on the scattering matrix and the dynamics of the motional wave function following detection of a first reflected photon ($\ket{\psi_r}$, Eq.~\eqref{eq:psir}). 
This approach neglects contributions to $\gr$ that arise from the anharmonicity $\sim \Delta$ of the J-C ladder between $0\rightarrow 1$ and $1\rightarrow 2$ excitations (represented by the red arrows in Fig.~\ref{fig:schemeLoss}(b)).
We will later show, by comparing with full master equation simulations, that the scattering matrix captures well important features of $\gr(t)$ and, in particular, oscillations due to strong optomechanical coupling. 
For our previous choice of detunings [(\ref{eq:optoml},\ref{eq:optdelc}), ensuring $\mathcal R(x_0)=0$], a central hole is imprinted in the conditional atomic motional state $\ket{\psi_r}$ (blue wave function in Fig.~\ref{fig:44MechanicalProjectionScatteringExpEv}(a)), reducing atomic population at positions where reflection is more unlikely. Note that some of the other experimental parameters ($\kappa_1/\kappa_2$ and $\eta$) have been changed relative to previous figures to make the relevant effects more visible.

To approximate $\gr(t)$, we consider the limit of a weakly driven cavity, such that the forces associated with the cavity field are negligible compared to the external trap. The subsequent dynamics of the atomic state are then dominated by the evolution purely in the trapping potential, $\ket{\psi_r(t)}=e^{-i\ham{trap}t}\ket{\psi_r}$ before further scattering events occur, as the atomic motion is highly isolated from its environment. 
Because of the overall mirror symmetry in $\ket{\psi_r}$ found for the discussed configuration, a revival of the wave function appears with periodicity in time $\pi/\omega_m$ as the atomic state evolves [red and green wave functions in Fig.~\ref{fig:44MechanicalProjectionScatteringExpEv}(a), see also the illustration in Fig.~\ref{fig:44MechanicalProjectionScatteringExpEv}(b)]. 

This time-evolving spatial distribution, combined with the sensitivity of the cavity response to the position of the atom, should result in a conditional time-dependent reflectance that manifests in $\gr(t)$ as,
\begin{equation}
\label{eq:g2}
    \gr( t)\approx \frac{\abs{S_r\ket{\psi_r( t)}}^2}{\mathcal R}\,,
\end{equation}
which compares the reflectance of the cavity at time $t$ after detection of a reflected photon to the initial reflectance $\mathcal R$ of the cavity, considering that intermediate scattering events are unlikely over the observation time. 

In Fig.~\ref{fig:44MechanicalProjectionScatteringExpEv}(c) (blue curve), we plot the predicted $\gr(t)$ from Eq.~\eqref{eq:g2}, for the spatial dynamics illustrated in Fig.~\ref{fig:44MechanicalProjectionScatteringExpEv}(a). We observe a bunching effect immediately after detection of the first reflected photon, as detection projects the atomic state into a configuration compatible with that event. The same cavity response is expected whenever the state revives, which for the symmetric configuration presented above, occurs with periodicity $\pi/\omega_m$.

To validate these results in the weakly driven regime, we have also performed a full master equation simulation of the driven system \eqref{eq:JCevolution} for a weak field input as described in Sec.~\ref{sec:heating} (orange dashed line in Fig. \ref{fig:44MechanicalProjectionScatteringExpEv}(c)). We observe good agreement with the results provided by Eq. \eqref{eq:g2} at times $t>1/\tilde\gamma$. 
At shorter times, we note an additional contribution to $\gr(t)$ further illustrated in the inset, which arises from the anharmonicity of the J-C ladder of a motionless atom and decays after a time $\sim 1/\tilde\gamma$.

Regarding the significance of these time-dependent oscillations in $\gr(t)$, we point out that they fundamentally differ from oscillations in reflection that could be observed, for example, by applying a classical momentum kick on the atom. In particular, in the latter case, given an atom originally in a stationary state (such as the motional ground state or a thermal state), an additional optical pulse (or a sudden variation in the trapping field) could induce motional oscillations in the atom. These would be already visible as temporal oscillations in the cavity output field $\ev{\ah\st{out}\dg(t) \ah\st{out}(t)}$, given a weak probe input. Note that these oscillations would only be significant if the kicking pulse contained many photons, given the small recoil energy of a single photon compared to the trapping frequency. In the presented scheme, however it is the detection of just a single photon which has an associated kick that is strong enough to induce significant oscillations, a key signature of strong optomechanical coupling. Furthermore, the conditional nature of this effect causes these oscillations to appear in the higher-order correlation of $\gr(t)$, rather than the unconditional reflectance itself.

Furthermore, the period of oscillations can be modified by tuning the driving frequency such that $\mathcal R(x_0+x\st{zp})=0$ (e.g. replacing $x_0\to x_0+x\st{zp}$ in Eqs. (\ref{eq:optoml}, \ref{eq:optdelc})). The detection of a reflected photon results in a conditional wave function whose probability amplitude is increased on one side of the trap, as illustrated in Fig. \ref{fig:44MechanicalProjectionScatteringExpEv}(d). After half a period, this state now oscillates to the opposite side of the trap (see Fig. \ref{fig:44MechanicalProjectionScatteringExpEv}(e)) which, in this configuration, manifests as antibunching ($\gr(\pi/\omega_m)<1$), restoring the natural periodicity $2\pi/\omega_m$ of the correlator $\gr(t)$, as we show in Fig. \ref{fig:44MechanicalProjectionScatteringExpEv}(f).

We now discuss the approximate conditions desired to observe large contrast in the time-dependent oscillations in $\gr(t)$. We begin by noting that our previous strong coupling conditions, based on achieving an effective length $\ell/x\st{zp}$ as small as possible (see Eq.~\eqref{eq:R}), do not directly translate into large oscillations in $\gr(t)$. In particular, as $\ell/x\st{zp}\rightarrow 0$, the atom is unlikely to be in a position around $x_0$ where the reflectance is suppressed, leading to $\mathcal R\rightarrow 1$. As a consequence, the small hole in the conditional wave function that occurs after detection cannot significantly increase the conditional reflectance, and thus $\gr(0)\rightarrow 1$.

To better interpret how intermediate situations may be optimal, one can explore a simplified uniform response in reflectance $\mathcal{R}(x)=1-(1-\mathcal{R}_0)\Theta\co{\ell-\abs{x}}$ for a homogeneous mechanical state $\abs{\psi(x)}^2=(2x\st{zp})^{-1} \Theta\co{x\st{zp}-\abs{x}}$; where $\Theta[x]$ denotes the step function that is $1$ for $x>0$ and 0 otherwise. In this toy model, one obtains that 
the maximum value of $\gr(0)$ occurs when $\ell(1+\mathcal{R}_0)\sim x\st{zp}$, which defines an optimal (non-zero) length for each choice of $\mathcal{R}_0$. The optimal configuration is a balanced cavity ($\mathcal{R}_0=0)$, where one would desire $\ell \sim x\st{zp}$.

To further illustrate this, in Fig. \ref{fig:71optimizingG2_k1k2}(a)
we calculate the scattering matrix approximation to $\gr(0)$~\eqref{eq:g2}, using the same parameters for $g_0$, $\gamma$, and $\kappa_2$ as in the experiment of Ref.~\cite{Samutpraphoot2019} and Fig.~\ref{fig:60MechanicalProjectionScattering_heatingScalingConditions_dat}(a). However, we now allow the Lamb-Dicke parameter (experimentally tunable through the intensity of the trapping potential) and the output port decay rate $\kappa_1$ to vary. Choosing for each set of $\eta$ and $\kappa_1$ the atom-cavity detuning that minimizes $\mathcal R(x_0)$, and driving in resonance with the dressed atomic frequency for an atom positioned at $x_0$, we heuristically observe that the largest values of $\gr(0)\sim 3$ appear in a region compatible with $\ell(1+\mathcal{R}_0)=6 x\st{zp}$ (red continuous line), which aligns with the intuition built from our toy model. 
In Fig. \ref{fig:71optimizingG2_k1k2}(b) we further show the overall variation of $\gr(t)$ along a full mechanical oscillation, $\Delta \gr\equiv \max_t \gr(t)-\min_t \gr(t)$, observing that the largest values $\Delta \gr\sim 2$ appear in a region compatible to those with larger $\gr(0)$. 

Here, one can also see a sharp change in $\gr(0)$ around $\kappa_1=\kappa_2$, that is more evident as $\eta<0.1$. In this latter regime, the effective atomic displacement over which the reflectance of the cavity varies $\ell/x\st{zp}\sim (\eta\sqrt{C\st{in}})^{-1}$ becomes much larger than the characteristic spread of the atomic state and, therefore, the response of the cavity becomes less sensitive to the position of the atom. For $\kappa_1/\kappa_2>1$, ensuring that no reflection occurs for an atom at the center of the trap $(\mathcal{R} _0=0)$ still allows to highly discriminate these central positions when a reflected photon is detected, which translates into a large  response $\gr(0)\sim 2$ even if the absolute variation of $\mathcal R(x)$ is very reduced. However, when the cavity leakage through the undetected channel exceeds the emission in the driven port $(\kappa_1\ll\kappa_2)$ there is no possible choice of atom-cavity detuning that allows for $\mathcal{R}_0$ to vanish (see Eq. \eqref{eq:optdelc}), and the detection of a reflected photon barely provides any information about the atomic position, which leads to the observed values $\gr(0)\sim 1$ when $\eta\ll 1$.

\subsection{Finite temperature}
\label{ap:finiteTemp}
In a real experimental situation, limitations in cooling or atom transport can prevent the atom from being prepared in its motional ground state, and instead the motional state might be given by a thermal density matrix at temperature $T$,
\begin{equation}
    \rh_T=\frac{e^{-\ham{trap}/(k_B T)}}{Z}\,,
\end{equation}
where $k_B$ is the Boltzmann constant and $Z=\tr{e^{-\ham{trap}/(k_B T)}}$ the partition function.
An important consequence is that its steady-state position uncertainty  $x_T\equiv \sqrt{\tr{\hat x^2\rh_T}}$ becomes temperature broadened as $x_T/x\st{zp}\approx \sqrt{2n\st{ph}+1}$, where $n\st{ph}\equiv \tr{\bh\dg\bh\rh_T}$ is the thermal phonon number that approximates $n\st{ph}\approx k_B T/\omega_m$ in the limit $n\st{ph}\gg1$. 

\begin{figure}[tbp]
\centering
\hspace*{-1cm}\includegraphics[width=0.45\linewidth]{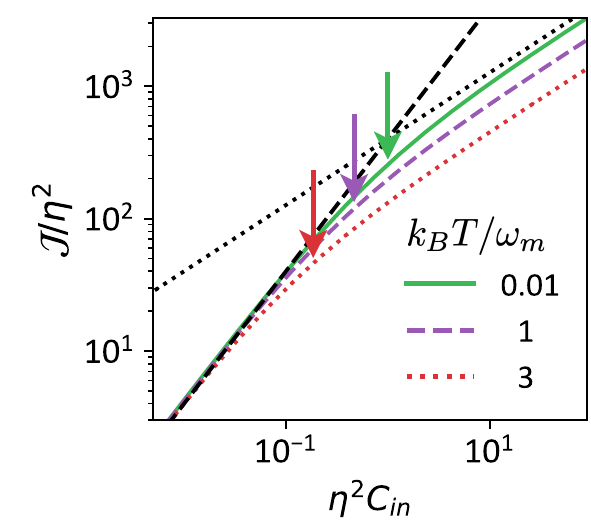}
  \caption{Average number of induced phonons per photon, normalized by the free-space expectation, $\mathcal{J}/\eta^2$, for an atom in a thermal state $\rho_T$ associated to temperatures $k_B T/\omega_m=0.01,\, 1,\,3$ (see legend), and increasing values of atom-cavity coupling. The rest of parameters are chosen as in Fig.  \ref{fig:44MechanicalProjectionScatteringExpEv}(a). We observe that the transition between the weak ($\mathcal J \propto \eta^2 C\st{in}$, black dashed line) and strong coupling limits ($\mathcal J \propto \eta \sqrt{C\st{in}}$, black dotted line) appears at the critical effective length $[\ell=x_T]$ marked with coloured arrows.}
 \label{fig:64thermalState}
\end{figure}

In analogy to the role played by $x\st{zp}$ in the zero-temperature limit, $x_T$ represents the characteristic temperature-dependent length of the system, and strong optomechanical coupling is expected to occur when $x_T\sim \ell$. To illustrate this, in Fig. \ref{fig:64thermalState} we calculate the average number of additionally induced phonons caused by a single photon as the intrinsic cooperativity increases, where each contribution can be obtained from the scattering matrix description as $\mathcal J_\alpha=\tr{\bh\dg\bh\, S_\alpha \rh_T S_\alpha\dg}$. Note that phonons already present in the thermal state now need to be subtracted from the number of phonons in the final conditional state, so that the net number of phonons added is $\mathcal J=\mathcal J_r+\mathcal J_t+\mathcal J_a-\tr{\bh\dg\bh\, \rh_T}$. Presenting the calculation for three different temperatures, we observe for each of them that $x_T=\ell$ (marked with arrows) defines the crossover, where $\mathcal{J}$ changes in scaling from $\eta^2 C\st{in}$ to $\eta \sqrt{C\st{in}}$. While this transition point occurs at smaller values of $\eta^2 C\st{in}$ as temperature is increased, the magnitude of the strong coupling effect also decreases, as evidenced by the decreased heating $\mathcal J/\eta^2$.

\begin{figure}[tbp]
\centering
\hspace*{-1cm}\includegraphics[width=0.6\linewidth]{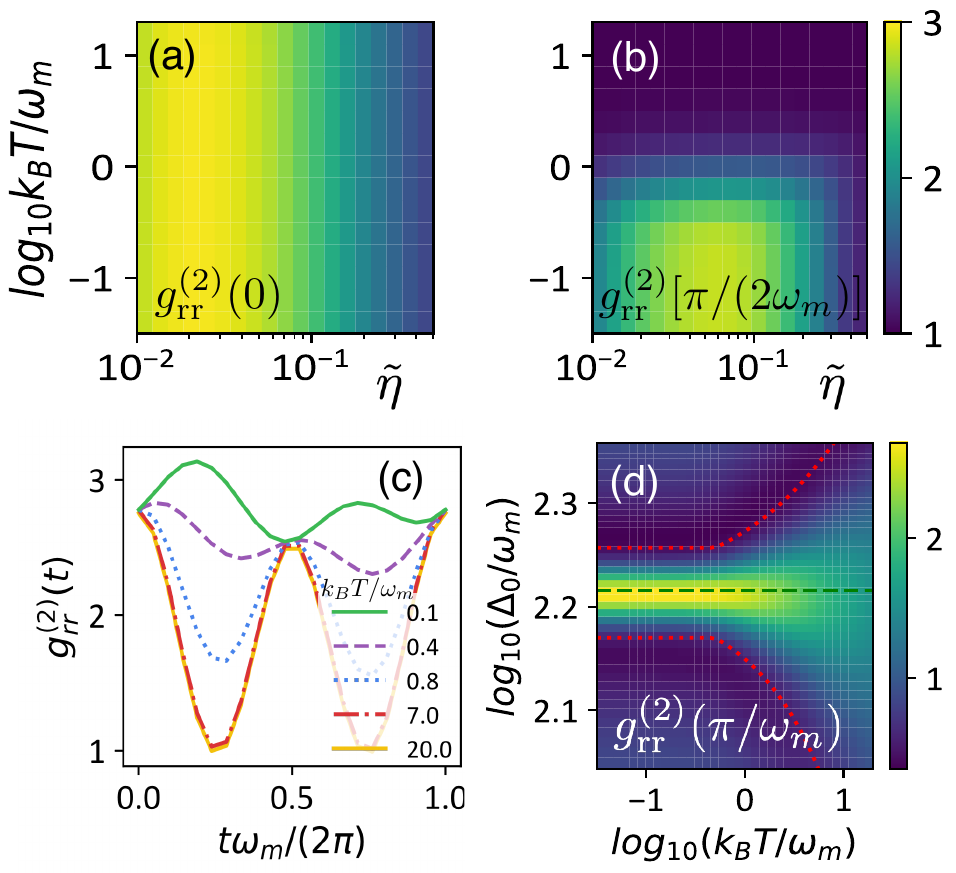}
  \caption{Value of the second-order time correlation function at times $\gr(0)$ (a) and $\gr[\pi/(2\omega_m)]$ (b), as a function of the temperature $T$ of the initial thermal state and the effective Lamb-Dicke parameter, $\tilde \eta=\eta x_T /x\st{zp}$, for fixed coupling ratio $\kappa_1/\kappa_2=1.6$ and the rest of parameters chosen as in Fig. \ref{fig:71optimizingG2_k1k2}.
  (c) Evolution of $\gr(t)$ associated to five different values of temperature for the choice of effective Lamb-Dicke parameter $\tilde \eta=0.05$. (d) Second-order time correlations half a mechanical oscillation away from initial detection, $\gr(\pi/\omega_m)$, for different driving frequencies and temperatures of the initial atomic thermal state, and fixed $\eta=0.05$. Bunching (light yellow) prevails for driving resonant with the atom-like frequency, $[\mathcal R(x_0)=0]$ (green-dashed line, Eq.~\eqref{eq:optoml}), while antibunching (dark blue) appears for a detuning compatible with $[\mathcal R(x_0\pm x_T)=0]$ (red dotted lines, see text). Calculations are performed within the scattering matrix model (see main text), observing convergence when allowing for up to 150 phonons in the explored range of temperatures. The rest of parameters are chosen as in Fig. \ref{fig:60MechanicalProjectionScattering_heatingScalingConditions_dat}.}
 \label{fig:64thermalState2}
\end{figure}

One can also analyze the effect that temperature has on the second-order correlations previously discussed. In Fig. \ref{fig:64thermalState2} we use the scattering matrix formalism to calculate $\gr(t)$ for different initial thermal states. For this, we assume that the dynamics of the thermal state conditioned to reflection of an initial photon, $\rh_{T,r}=S_r\rh_T S_r\dg/\tr{S_r\rh_T S_r\dg}$, is dominated by the evolution purely in the trapping potential, $\rh(t)=e^{-i\ham{trap}t}\rh e^{i\ham{trap}t}$, so that one can approximate the second-order correlator as $\gr(t)\approx \tr{S_r\rh_{T,r}(t) S_r\dg}/\tr{S_r\rh_T S_r\dg}$. 

The fact that the characteristic width of the state, $x_T$, now depends on temperature yields some interesting phenomena. 
First, although Eq.~\eqref{eq:g2} was formally derived assuming a pure initial state, one can see that $\gr(0)$ only in fact depends on the position probability distribution, suggesting that the purity of the state is irrelevant.
Thus, one might expect $\gr(0)$ to be independent of temperature, as long as the trapping frequency is adjusted so that the effective Lamb-Dicke parameter $\tilde \eta= \eta x_T/x\st{zp}$ remains constant. This independence is illustrated in  Fig.~\ref{fig:64thermalState2}(a), where we plot $\gr(0)$ as a function of temperature and $\tilde{\eta}$. 

However, the thermal nature of the state is expected to play a role in the subsequent dynamics of $\gr(t)$. To investigate this, in Fig.~\ref{fig:64thermalState2}(b) we show the second-order correlation function associated to time $t'=\pi/(2\omega_m)$ (the two reflected photons are separated by a quarter of the mechanical oscillation period). At this time delay, we see that $\gr(t')$ does retain a temperature dependence for fixed $\tilde{\eta}$, and tends toward 1 at larger temperatures. 

To understand this, note that following the detection of the first reflected photon, evolution under $\ham{trap}$ during a time $t'=\pi/(2\omega_m)$ causes the position quadrature to fully transform to momentum, and vice versa. In particular, the spatial width of this conditioned state at time $t'$, $\tr{\hat x^2\rh_{T,r}(t')}/x\st{zp}^2$, equals the width in momenta at initial time, $\tr{\hat p^2\rh_{T,r}(0)}/p\st{zp}^2$, where $p\st{zp}\equiv 1/(2x\st{zp})$. 
On the other hand, the resulting increase in kinetic energy due to the momentum imparted by detection for fixed $\tilde{\eta}$ is given by $\sim \ell^{-2}\sim n\st{ph}^{-1}$. This becomes negligible compared to the kinetic energy of the thermal state $\sim n\st{ph}$.
It then follows that $\tr{\hat p^2\rh_{T,r}(0)}\approx \tr{\hat p^2\rh_T}$ in the regime $n\st{ph}\gg 1$, which leads to the observed limit $\gr(t')\to 1$.
Thus, by fixing $\tilde{\eta}$, the overall variation of $\gr(t)$ in the limit $n\st{ph}\gg 1$ then oscillates between the result $\gr(0)$ also expected at zero temperature, to the value $\gr(t')\sim 1$ appearing  for $n\st{ph}\gg 1$ after one quarter of the mechanical period, as we plot in Fig.~\ref{fig:64thermalState2}(c). As a consequence, higher temperature can in fact lead to greater contrast in the temporal oscillations of the second-order correlation function.

Finally, in Fig.~\ref{fig:64thermalState2}(d), we plot $\gr(\pi/\omega_m)$ as a function of atom-laser detuning and temperature. Here, we fix the trapping frequency $\omega_m$ to yield a (zero-temperature) Lamb-Dicke parameter of $\eta=0.05$. We observe that after half a mechanical oscillation, bunching and antibunching occur for driving frequencies that satisfy the conditions $\mathcal{R}(x_0)=0$ (green dashed line) and $\mathcal{R}(x_0\pm x_T)=0$ (red dotted line), respectively. The latter condition now depends on the temperature of the state through the temperature-broadened $x_T$, so that driving frequencies associated to bunching and antibunching separate as the temperature increases. Taking now the limit of large temperature, we observe that second-order correlations tend to the Poissonian result $\gr=1$ as the change in the wave function, restricted to a region $\ell$, is less significant when the temperature increases and the atom is more spread.
Still, we note that deviations from the Poissonian result $\gr=1$ of order $40\%$ can be readily observable for phononic occupations in the order of $n\st{ph}\approx 7$, compatible with Ref. \cite{Samutpraphoot2019}.

\subsection{Conclusions and outlook}
Taking advantage of the narrow linewidth of a single atom, we have shown that it is possible to reach the single-photon strong coupling regime of optomechanics, even when the cavity linewidth is prohibitively large. We have shown that this optomechanical strong coupling can give rise to anomalously large motional heating, and to motionally-induced oscillations in the second-order correlation function of the light reflected from the cavity.

From the perspective of utilizing atom-cavity systems to realize coherent spin-photon interfaces, such as for quantum information processing, our work shows that there is the possibility to get strongly entangled with other undesired degrees of freedom, in the form of phonons. It is therefore important to specifically account for this effect when analyzing and optimizing protocols, especially in systems with high cooperativity and large spatial variations of the vacuum Rabi splitting $g(x)$. On the other hand, such a platform would be unique in enabling the study of quantum optomechanics in the strong coupling regime. For example, it would be interesting to investigate how to exploit such systems to realize strongly non-Gaussian dynamics. Separately, with the possibility to scale atom-cavity interfaces to multiple atoms and/or cavities~\cite{Samutpraphoot2019}, it might be possible to observe interesting strongly correlated optomechanical states at the many-body level~\cite{Manzoni2017a}.

\subsection*{Acknowledgements}
The authors thank R. Bettles, P. Samutpraphoot, T. \DJ{}or\dj{}evi\ifmmode \acute{c}\else \'{c}\fi{}, P. Ocola, H. Bernien and B. Grinkemeyer for insightful discussions. We acknowledge support from the European Union’s Horizon 2020 research and innovation programme, under European Research Council grant agreement No 101002107 (NEWSPIN), the Government of Spain (Europa Excelencia program EUR2020-112155, Severo Ochoa program CEX2019-000910-S, and MICINN Plan Nacional Grant PGC2018-096844-B-I00), Generalitat de Catalunya through the CERCA program, AGAUR Project No. 2017-SGR-1334, 
Fundaci\'o Privada Cellex, Fundaci\'o Mir-Puig, 
and Secretaria d’Universitats i Recerca del Departament d’Empresa i
Coneixement de la Generalitat de Catalunya, co-funded
by the European Union Regional Development Fund
within the ERDF Operational Program of Catalunya
(project QuantumCat, ref. 001-P-001644).
J.A.-L. acknowledges support from 'la Caixa' Foundation (ID 100010434) through the fellowship LCF/BQ/ES18/11670016.

\appendix
\section{Scattering approach for a single atom}
\label{ap:scattering}
In Eq.~\eqref{eq:reflFactorT}, we have provided the relation between the reflected and input fields for a coherently, weakly driven cavity and when the atom is placed in a fixed position~\cite{Gardiner1985}.
Here, we will show that Eq.~\eqref{eq:reflFactorT} relates to the scattering matrix for an atom whose motion constitutes a dynamical degree of freedom~\cite{Xu2015,caneva2015quantum}.
In particular, we consider a situation where the joint cavity-atom-motional system is in its ground state $\ket{0_c\downarrow}\ket{\psi_0}$, and a single monochromatic photon of frequency $\omega_l$~(or corresponding detuning $\Delta_0$) is sent in. The S-matrix formally provides the transformation from the total input state $\ket{\Psi\st{in}}=\ket{0_c\downarrow}\ket{\psi_0}\ket{1\st{in}}$, to the output state at infinite time, $\ket{\Psi\st{out}}=S\ket{\Psi\st{in}}$. For the output, we take into account that the input photon could have been emitted through a detectable reflection channel with rate $\kappa_1$ ($\ket{1_r}$), undetectable cavity channels (transmission or loss) with rate $\kappa_2$ ($\ket{1_t}$), or spontaneously decayed with rate $\gamma$ after exciting the atom ($\ket{1_a}$). 
As the total energy of the system is conserved, the frequency of the output photon carries information about any possible change in the phononic state of the atom. To suppress any effect arising from this additional entanglement, we consider that the implemented detection scheme is not frequency-resolving. We also focus on the unresolved sideband regime, $\omega_m\ll \gamma,\kappa$, where the characteristic time-scale of the optomechanical interaction between the cavity and the internal state of the atom is much faster than the atomic dynamics inside the mechanical trap, which allows to describe the response of the cavity as effectively diagonal in the atomic position basis.

For an atom fixed in position $x$, the input photon then scatters as,
\begin{equation}
\label{eq:apScatOut}
     S(x)\ket{1\st{in}}
     =S_r(x)\ket{1_r}+S_t(x)\ket{1_t} +S_a(x)\ket{1_a}\,,
\end{equation}
which obey the relation $\abs{S_r(x)}^2 +\abs{S_t(x)}^2 +\abs{S_a(x)}^2=1$ to conserve the norm of the scattered state. To calculate these components, each of the matrix elements can be expressed in terms of the eigenvectors, $\ket{\beta(x)}$, and eigenvalues, $\lambda_\beta(x)$, of the effective (non-Hermitian) atom-cavity Hamiltonian $\hat H\st{eff}(x)=-(\Delta+\Delta_0+i\kappa/2) \ah\dg \ah  -(\Delta_0+i\gamma/2) \sh\dg \sh + g(x)\pa{\sh \ah\dg +\sh\dg \ah}$ as,
\begin{equation}
\label{eq:defSr}
    \begin{split}
        S_r(x) &=1+i\kappa_1\sum_\beta\braket{1_c\downarrow}{\beta(x)}\frac{1}{\lambda_\beta(x)}\braket{\beta^*(x)}{1_c\downarrow}\,, \\
        S_t(x) &=i\sqrt{\kappa_1 \kappa_2} \sum_\beta\braket{1_c\downarrow}{\beta(x)}\frac{1}{\lambda_\beta(x)}\braket{\beta^*(x)}{1_c\downarrow}\,, \\
        S_a(x) &=i\sqrt{\kappa_1\gamma}e^{ik_c x}\sum_\beta\braket{0_c\uparrow}{\beta(x)}\frac{1}{\lambda_\beta(x)}\braket{\beta^*(x)}{1_c\downarrow}\,, \\
    \end{split}
\end{equation}
 which requires the orthogonality relation $\braket{\beta^*(x)}{\beta(x)}=1$ due to the non-Hermitian nature of the Hamiltonian. To calculate these scattering elements, one can note that this Hamiltonian is block-diagonal and, given the single-photon input, one can restrict to the relevant subspace spanned by states $\ket{{0_c,\uparrow}}$ $\ket{1_c,\downarrow}$. Projecting there, one gets,

\begin{equation}
\begin{split}
\label{eq:sigss}
 S_r(x) & =1-\frac{i\kappa_1}{\Delta_0+\Delta+i\kappa/2-\frac{g(x)^2}{\Delta_0+i\gamma/2} }\,, \\
    S_t(x) &= \frac{-i\sqrt{\kappa_1 \kappa_2}}{\Delta_0+\Delta+i\kappa/2-\frac{g(x)^2}{\Delta_0+i\gamma/2} }\,, \\
    S_a(x) &=\frac{-i\sqrt{\kappa_1\gamma}}{\Delta_0+\Delta+i\kappa/2-\frac{g(x)^2}{\Delta_0+i\gamma/2} }\frac{g(x) e^{ik_c x}}{\Delta_0+i\gamma/2}\,.
\end{split}
\end{equation}
This leads to the scattering elements $S_\alpha(x)$ associated to the different possible routes the input photon could eventually decay through, $\alpha\in \lla{r,t,a}$. A more rigorous derivation based on Nakajima-Zwanzig leading to an analogous result can be found in ~\cite{Neumeier2018}.

If one now considers the atom to be initially in a superposition of different locations inside the cavity, $\ket{\psi_0}=\int dx\,  \psi_0(x)\ket{x}$, the probability of decaying in either of the emission channels is given by, $\abs{S_\alpha\ket{\psi_0}}^2=\int dx \abs{S_\alpha(x)}^2\abs{\psi_0(x)}^2$ and, if a photon $\ket{1_\alpha}$ was detected in either of these channels, the measurement then projects the atomic wave function into the conditional state, $\ket{\psi_\alpha}=S_\alpha\ket{\psi_0}/\abs{S_\alpha\ket{\psi_0}}$.

To get a better intuition on how the scattering matrices depend on atomic position, one can investigate these equations in the limit $C\st{in}\gg 1$ for the optimal choice of parameters motivated in the main text~(\ref{eq:optoml},\ref{eq:optdelc}), and $\kappa_1=2\kappa_2$. Then, expanding $S_{\alpha}$ to linear order in $\delta x=x-x_0$, one finds,
\begin{equation}
\begin{split}
\label{eq:delsig}
 S_r(x)&\approx -i\frac{\eta \sqrt{C\st{in}}}{\sqrt 2}\frac{\delta x}{x\st{zp}}\,, \\
 S_t(x)&\approx -\frac{1}{\sqrt 2}\pa{1+i\frac{\eta \sqrt{C\st{in}}}{\sqrt 2}\frac{\delta x}{x\st{zp}}}\,, \\
 S_a(x) &\approx \frac{1}{\sqrt 2}\co{1+i\pa{\frac{\eta \sqrt{C\st{in}}}{\sqrt 2}+\eta}\frac{\delta x}{x\st{zp}}} \,,
\end{split}
\end{equation}
which are valid for $\abs{\delta x}\ll \ell$. In the opposite limit, $\abs{\delta x}\gg \ell$, the cavity becomes out of resonance and one recovers perfect reflectance: $\abs{S_r(x)}\to 1$ and $\abs{S_t(x)},\abs{S_a(x)}  \to 0$. 

\end{document}